# A novel route to phase formation of cobalt oxyhydrates using KMnO$_4$ as an oxidizing agent

C.-J. Liu,[a,*] C.-Y. Liao,[a] L.-C. Huang,[a] C.-H. Su,[a] S. Neeleshwar,[b] and Y.-Y. Chen[b]

[a]Department of Physics, National Changhua University of Education, Changhua, 500 Taiwan, R. O. C.

[b]Institute of Physics, Academia Sinica, Taipei, Taiwan, R. O.C.

*To whom correspondence should be addressed. Phone: +886-4-723-2105 ext-3337.   E-mail: liucj@cc.ncue.edu.tw.

ABSTRACT

We have first successfully synthesized the sodium cobalt oxyhydrate superconductors using KMnO$_4$ as a de-intercalating and oxidizing agent.  It is a novel route to form the superconductive phase of Na$_x$CoO$_2$·yH$_2$O without  resort to the commonly used Br$_2$/CH$_3$CN solution.  The role of KMnO$_4$ is to de-intercalate the Na$^+$ from the parent compound Na$_{0.7}$CoO$_2$ and oxidize the Co ion as a result.  The higher molar ratio of KMnO$_4$ relative to the sodium content tends to remove more Na$^+$ from the parent compound and results in a slight expansion of the *c*-axis in the unit cell.  The superconducting transition temperature is 4.6 - 3.8 K for samples treated by aqueous KMnO$_4$ solution with the molar ratio of KMnO$_4$ relative to the sodium content in the range of 0.03  and 2.29.

# 1. Introduction

The recent discovery of superconductivity [1] in hexagonal cobalt oxyhydrates $Na_xCoO_2 \cdot yH_2O$ with $T_c = 4 - 5$ K has attracted great attention [2-7] because it is unique to have the water molecule intercalated into the parent structure of $Na_{0.7}CoO_2$ which is crucial for the occurrence of superconductivity. It is also a particularly interesting system for comparison with the high-$T_c$ cuprates in terms of the structure-electronic state correlations in view of the fact that both have 2D layers (triangular $CoO_2$ layers and square $CuO_2$ layers) in structure and have spin 1/2 ions ($t_{2g}^5$ for $Co^{4+}$ in low spin state and $t_{2g}^6 e_g^3$ for $Cu^{2+}$) in electron configuration. Besides, this family of materials is of particular interest because of their magnetic and thermoelectric properties. The parent compound $Na_xCoO_2$ is a potential candidate for thermoelectric applications due to its high electrical conductivity, large thermopower and low thermal conductivity [8]. There exists four phases [9.10] in the related cobalt oxides: (1) α-$Na_xCoO_2$ ($0.9 \leq x \leq 1$, O3 phase); (2) α'-$Na_{0.75}CoO_2$ (O'3 phase); (3) β-$Na_xCoO_2$ ($0.55 \leq x \leq 0.6$, P3 phase); and (4) γ-$Na_xCo_yO_2$ ($0.55 \leq x/y \leq 0.74$, P2 phase). The O3 phase represents 3 $CoO_2$ layers in the unit cell with the $Na^+$ in the octahedral surroundings, while P2 phase represents 2 $CoO_2$ layers in the unit cell with the $Na^+$ in the trigonal prismatic surroundings. The O'3 phase is the monoclinic distortion of O3 phase.

Takada et al. obtained the superconducting phase of $Na_xCoO_2 \cdot yH_2O$ by immersing $Na_{0.7}CoO_2$ powders in $Br_2/CH_3CN$ solution followed by filtering and rinsing [1]. Park et al. [11] reported the alternative route to make the superconductive $Na_xCoO_2 \cdot yH_2O$. This process is generally considered as a chemical oxidation by removing Na partially before the $H_2O$ is intercalated between the $CoO_2$ layers and Na layers. The purpose of this paper is to report a novel route of preparing the superconducting $Na_xCoO_2 \cdot yH_2O$ phase using aqueous $KMnO_4$ solution as a de-intercalating and oxidizing agent. It is an alternative route to use oxidant rather than $Br_2/CH_3CN$ solution to form the superconducting phase of $Na_xCoO_2 \cdot yH_2O$.

# 2. Experimental

The superconducting cobalt oxyhydrates $Na_xCoO_2 \cdot yH_2O$ were prepared following the procedures below: (1) Preparation of parent material $Na_{0.7}CoO_2$. Polycrystalline powders of $Na_{0.7}CoO_2$ were synthesized by quantitatively mixing high-purity powders of $Na_2CO_3$ and CoO. The mixed powders were ground thoroughly using a Retch MM2000 laboratory mixer mill, followed by a rapid heat-up procedure [12] at 700°C in order to avoid the loss of Na in the heating process. (2) De-intercalation of Na and Oxidation. The resulting powders (0.5 -1 g) were immersed and stirred in 50 ml of water solution with different molar ratios of $KMnO_4$/Na labeled as 0.05X – 2.29X at room temperature for 5 days. (3) Formation of superconductive phase. The products were carefully filtered and washed thoroughly with deionized water, followed by drying at ambient conditions for 10 - 20 hrs in order to remove the powders from the filter paper. The dried powders were then stored in a wet chamber with sufficient humidity for further structural and magnetic characterization. Powder x-ray diffraction (XRD) patterns were obtained using a. Shimadzu XRD-6000 diffractometer equipped with Fe Kα radiation. The sodium content was determined by using a Perkin Elmer Optima 3000 DC inductively coupled plasma - atomic emission spectrometer (ICP-AES). Before the chemical analysis, samples are dehydrated by heating at 300℃ in air for 12 h. Thermogravimetric analysis (TGA) was carried out by using a Perkin Elmer Pyris 1 thermogravimetric analyzer. A commercial SQUID magnetometer (Quantum Design) was used to characterize the superconducting transition temperature of the samples.

**3. Results and discussion**

Fig. 1 shows the powder x-ray diffraction patterns (XRD) of the products obtained by different molar ratios of $KMnO_4$ relative to Na content. The XRD patterns for ≤ 0.1X samples are a mixture of a fully-hydrated superconducting phase and a non-superconducting dehydrated phase, which is similar to bromine-treated samples with substoichiometric or stoichiometric $Br_2/CH_3CN$ solutions [2]. Single phase of fully hydrated $Na_xCoO_2 \cdot yH_2O$ occurs for $KMnO_4$/Na = 0.3. For 0.5 ≤ $KMnO_4$/Na ≤ 2.29, there is a very tiny peak appearing at 2θ ≈ 16°, the characteristic (002) peak of so-called y = 0.6 intermediate hydrated phase [5], in addition to the diffraction peaks of the fully hydrated phase. The sodium contents are determined by using inductively coupled plasma atomic emission spectroscopy (ICP-AES) and shown in Table I. Chemical analyses show that the sodium contents in $Na_xCoO_2 \cdot yH_2O$ systematically decreases with increasing molar ratio of $KMnO_4$/Na. The values of x are 0.38, 0.32, 0.28 for 0.3X, 0.5X, and 2.29X samples, respectively. These results confirm that the role of $KMnO_4$ is acting as an oxidizing agent to partially de-intercalate the Na from the structure and hence oxidize the electronically active $CoO_2$ layers. The *c*-axis of the unit cell in $Na_xCoO_2 \cdot yH_2O$ tends to increase with increasing molar ratio of $KMnO_4$/Na from19.669 Å for the 0.3X sample to 19.735 Å for the 2.29X sample but with little changes in the *a*-axis

The thermal stability and water content of 0.3X sample are checked and determined by heating the sample in flowing $O_2$ at the slowest rate of 0.1℃/min available to the Perkin Elmer Pyris 1 thermogravimetric analyzer (TGA). Fig. 2 indicates a multi-stage loss of water with relatively unstable intermediates [13], being consistent with the thermally unstable nature of the fully hydrated phase [5,14]. The water content of fully hydrated phase is estimated to contain 1.45 and 1.55 $H_2O$ pre formula unit by taking the weight loss at 320 ℃ and 600℃, respectively, as the fully dehydrated phase (y = 0), assuming no oxygen deficiency in the sample for the present estimations.

Fig. 3 shows the zero-field cooled and field cooled magnetization data of 0.3X, 0.5X, and 2.29X samples measured in a dc field of 10 Oe. The onset superconducting transition is observed at about 4.6 K, 4.5 K, and 3.8 K for 0.3X, 0.5X, and 2.29X, respectively. The mass magnetization at 1.8 K is -$1.28 \times 10^{-2}$ emu/g in the zero-field cooling measurements, which is approximately 31 % of the theoretical value for perfect diamagnetism.

## 4. Conclusions

We have first synthesized the superconductive cobalt oxyhydrates $Na_xCoO_2 \cdot yH_2O$ using $KMnO_4$ as an oxidizing agent instead of using $Br_2/CH_3CN$ solutions. The role of $KMnO_4$ is to de-intercalate the Na from the structure and hence oxidize the Co ion based on the electron neutrality. The higher molar ratio of $KMnO_4$ relative to Na content used to treat the samples leads to more removal of Na. The superconductive phase of $Na_xCoO_2 \cdot yH_2O$ is commonly obtained by $Br_2/CH_3CN$ solutions, which is highly toxic by ingestion and inhalation. This new route might also indicate that $KMnO_4$ has the potential to treat other layered oxide materials with similar function of de-intercalation and oxidation and to have mass production of superconducting samples.

**Acknowledgment.** This work is supported by National Science Council of ROC under the grant No. NSC 92-2112-M-018-005.

**References**

(1) K. Takada, H., Sakurai, E. Takayama-Muromachi, F. Izumi, R.A. Dilanian, T. Sasaki, Nature 422 (2003) 53.

(2) R. E. Schaak, T. Klimczuk, M. L. Foo, R. J. Cava, Nature 424 (2003) 527.

(3) J. Cmaidalka, A. Baikalov, Y. Y. Xue, R. L. Meng, C. W. Chu, Physica C 403 (2004) 125.

(4) B. G. Ueland, P. Schiffer, R.E. Schaak, M. L. Foo, V. L. Miller, R. J. Cava, Physica C 402 (2004) 27.

(5) M. L. Foo, R.E. Schaak, V. L. Miller, T. Klimczuk, N. S. Rogado, Y. Wang, G. C. Lau, C. Craley, H. W. Zandbergen, N. P. Ong, R. J. Cava, Solid State Commun. 127 (2003) 33.

(6) J. D. Jorgensen, M. Avdeev, D.G. Hinks, J. C. Burley, S. Short, cond-mat/0307627 (2003).

(7) H. D.Yang, J.-Y Lin, C. P. Sun, Y. C. Kang, K. Takada, T. Sasaki, H. Sakurai, E. Takayama-Muromachi, cond-mat/0308031 (2003).

(8) I. Terasaki, Y. Sasago, K. Uchinokura, Phys. Rev. B 56 (1997) R12685.

(9) C. Fouassier, G. Matejka, J.-M. Reau, P. Hagenmuller, J. Solid State Chem. 6 (1973) 532.

(10) M. Mikami, M. Yoshimura, Y. Mori, T. Sasaki, Jpn. J. Appl. Phys. 41 (2002) L777.

(11) S. Par, Y. Lee, A. Moodenbaugh, Vogt, T. Phys. Rev. B 68 (2003) 180505.

(12) T. Motohashi, E. Naujalis, R. Ueda, K. Isawa, M. Karppinen, H. Yamauchi, Appl. Phys. Lett. 79 (2001) 1480.

(13) M. E.Brown, Introduction to Thermal Analysis, Chapman and Hall, New York, 1988.

(14) D. P. Chen, H. C. Chen, A. Maljuk, A. Kulakov, H. Zhang, C. T. Lin, cond-mat/0401636 (2004).

Table I. Sodium content and lattice constants of $Na_xCoO_2 \cdot yH_2O$ prepared using $KMnO_4$ as oxidant

| Molar ratio of $KMnO_4$/Na | Sodium content $x^a$ | $a$ axis[b] Å | $c$ axis[b] Å |
|---|---|---|---|
| 0.3X | 0.38 | 2.8249(1) | 19.669(1) |
| 0.5X | 0.32 | 2.8248(1) | 19.679(1) |
| 2.29X | 0.28 | 2.8250(2) | 19.735(2) |

[a]The error in weight % of each element in ICP-AES analysis is ±3 %, which corresponds to an estimated error of ±0.02 per formula unit.

[b]Lattic constants are determined by least squares refinement using the XRD data between 2θ of 5° and 90° based on a hexagonal lattice with space group $P6_3/mmc$.

Figure and Captions

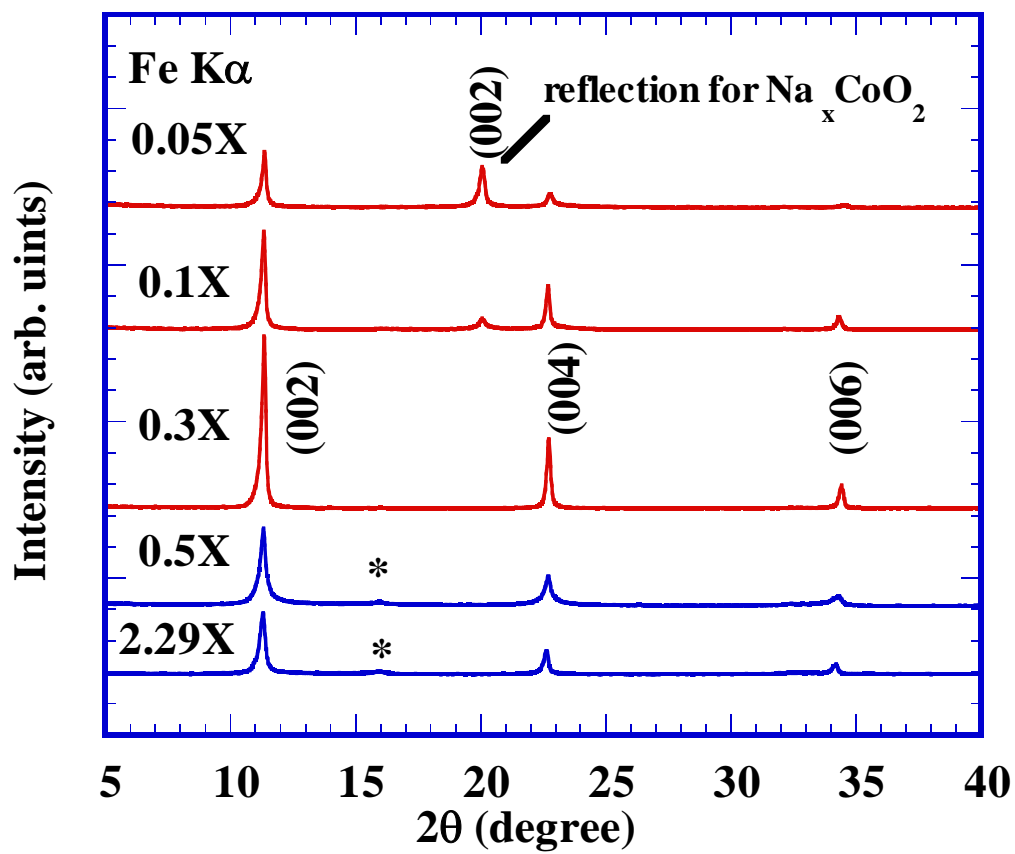

Fig. 1. Powder x-ray diffraction (XRD) patterns for $Na_xCoO_2 \cdot yH_2O$ prepared using different molar ratios of $KMnO_4$ relative to Na content. The 0.3 X and 2.29 X represent the molar ratios of $KMnO_4$ relative to Na content are 0.3 and 2.29, respectively. The asterisk indicates the tiny peak for the (002) diffraction peak for the so-called $y = 0.6$ intermediate hydrated phase.

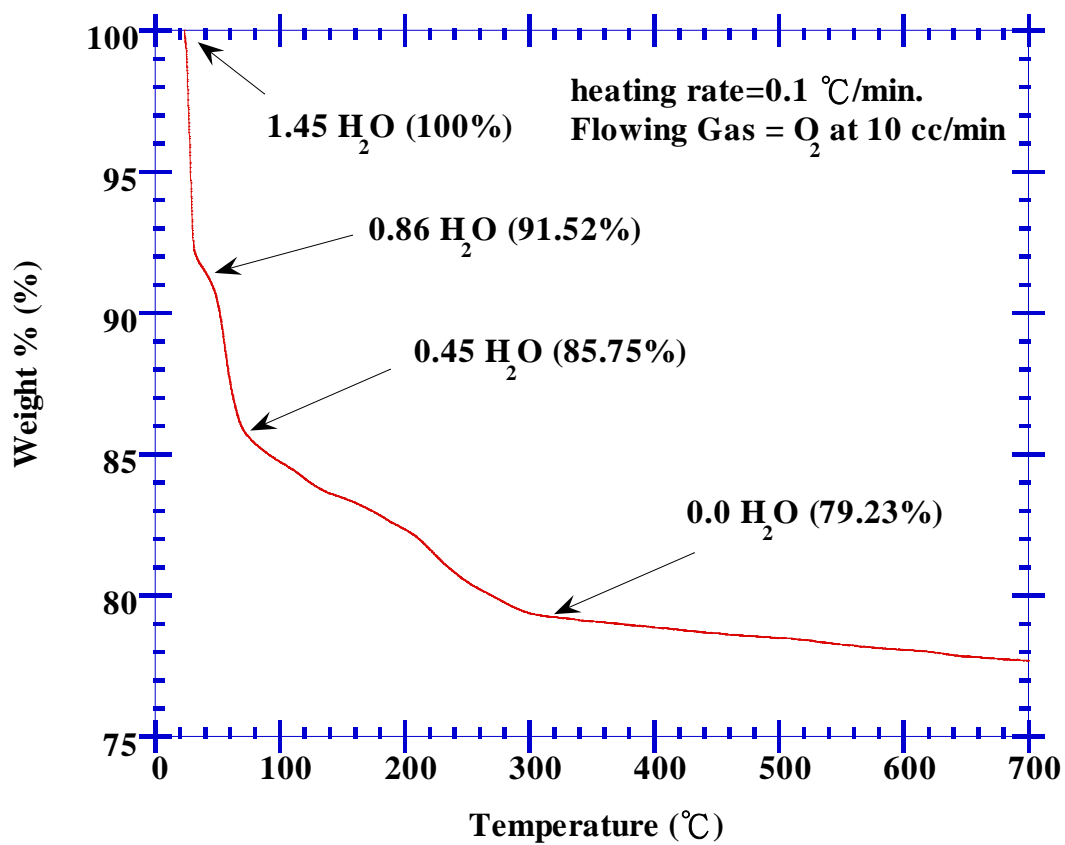

Fig. 2. Thermogravimetric analysis of $Na_xCoO_2 \cdot yH_2O$ (0.3X) with a heating rate of 0.1℃/min in flowing oxygen. The water content is determined by assuming a complete loss of water at 320℃.

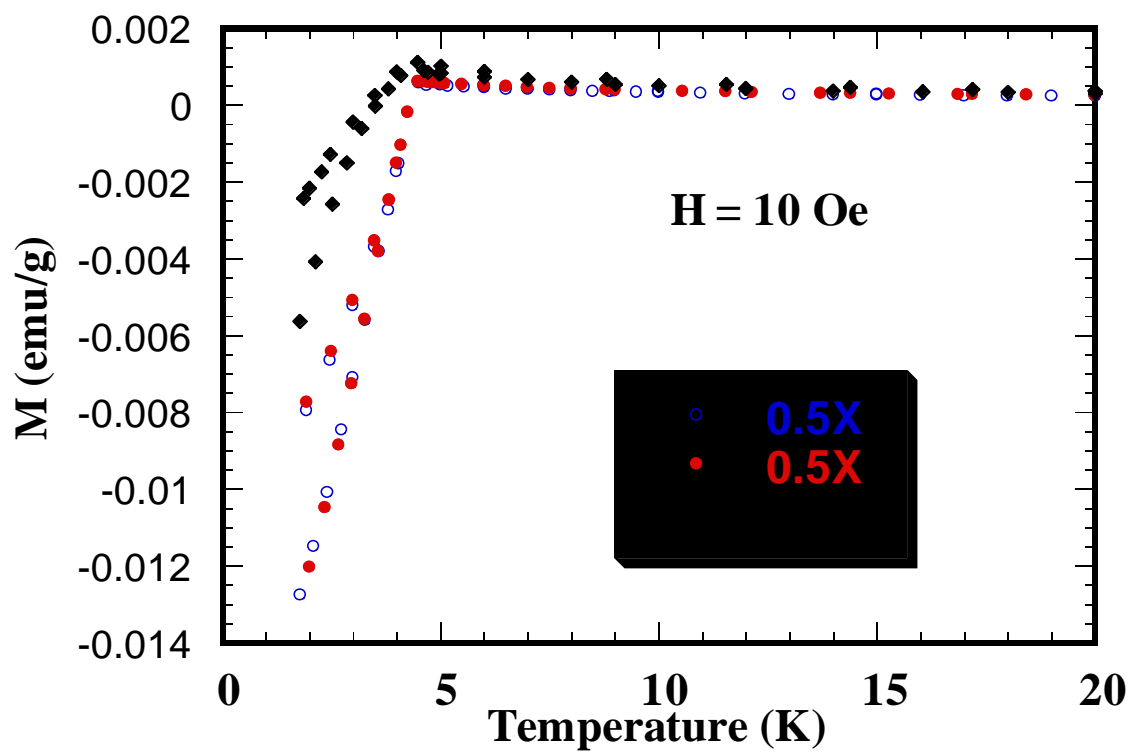

Fig. 3. Zero-field cooled and field cooled d.c. magnetization for $Na_xCoO_2 \cdot yH_2O$ (solid circle: 0.3X; open circle: 0.5X; solid diamond: 2.29X).